\documentclass[11pt,a4paper]{article}
\usepackage{amsfonts}
\usepackage{graphicx}
\usepackage{amsmath}
\usepackage{hyperref}
\usepackage{enumerate}
\usepackage{amsmath,amssymb}
\usepackage{slashed,mathrsfs}
\usepackage{feynmp,subfigure}
\usepackage{verbatim,graphicx}
\usepackage[sub,ovp]{psfragx}
\usepackage{overpic}

\RequirePackage{mathrsfs} \RequirePackage[sc]{mathpazo}
\RequirePackage{wasysym} \RequirePackage{setspace}
\newtheorem{theorem}{Theorem}
\newtheorem{acknowledgement}[theorem]{Acknowledgement}

\newcommand{\be}{\begin{equation}}
\newcommand{\ee}{\end{equation}}
\newcommand{\bea}{\begin{eqnarray}}
\newcommand{\eea}{\end{eqnarray}}
\input{tcilatex}
\begin{document}

\date{}
\title{ \rightline{\mbox{\small
{LPHE-MS/10-2022}}}\textbf{Massive Dark Photons as Hot Dark Matter}}
\author{Salah Eddine Ennadifi$^{1}$\thanks{%
ennadifis@gmail.com} \\
%EndAName
\\
{\small $^{1}$LHEP-MS, Faculty of Sciences, Mohamed V University, Rabat,
Morocco }\\
}
\maketitle

\begin{abstract}
Motivated by the growing interest in the existence of new massive gauge
bosons, we suggest that massive dark photons $A^{\prime }$ can be a
consequence of a broken new abelian symmetry $U(1)_{X}^{\prime }$. Such a
dark symmetry $U(1)_{D}^{\prime }$ is afterwards supposed to be associated
with the conservation of the weak number $W$ belonging to the Weakling
Interacting Slim Particles $U(1)_{W}$, being a strong candidate for Hot Dark
Matter. The latters correspond then to light dark photons $m_{A}^{\prime }$ $%
\lesssim $ $keV$ from a weak symmetry breaking scale $\gtrsim $ $10keV$.

\end{abstract}

\newpage

\section{Introduction}

During the last decades, a lot of theoretical speculations as well as
experimental searches have been driven towards the existence of new
particles beyond the Standard Model (SM) of particle physics \cite{1,2,3,4}.
However, due to the negative results of all these searches and hence the
increasing frustration for the failure to discover any of these hypothetical
new particles, these attempts have been increasingly challenged. As the
prospect of a breakthrough along these routes is dwinkling, the interest in
a hidden sector---hidden because neutral under the SM gauge symmetries---is
mounting: Perhaps no new particles have been observed merely because they do
not interact via the SM gauge interactions \cite{5,6,7,8,9}.

The hidden sector is supposed to exist as a solution to the problem of the
dark side of the universe whose presence is motivated by gravitational
physics \cite{10,11}, where the hidden particles serve as possible
candidates for the dark sector \cite{6,12,13}. Particularly, depending on
the model, the hidden sector may contain few or many states, and these can
be fermions, bosons or both; and whose the existence is believed needful to
account for astrophysical data \cite{5,6,7}. Additionally, these hidden
particles can communicate with the visible sector. In particular, other than
their gravitational interaction, they can communicate through their putative
weak interaction, because otherwise, there would be little hope for
detecting experimentally particles belonging to the hidden sector as in the
direct- and indirect-detection searches of weakly interacting massive
particles \cite{14,15}. For the same purpose, we must focus our prospects on
supposing that the hidden and the visible sectors also interact by means of
a narrow window, i.e., a portal, in a way that although is weak, it can be
(at least in principle) experimentally attainable \cite{16,17,18,19,20}.

Among the various candidates for the Dark Matter (DM), the Dark Photon (DP)
is a popular candidate for the very light DM particles that have been
largely dealt with \cite{21,22,23}. The DP constitutes a typical example of
a spin-1 bosonic light DM. It is associated with the minimal extention of
the SM gauge symmetry and is also well-motivated by major esteems \cite%
{24,25}. Similarly to other candidates, i.e., QCD axions, and axion-like
particles (ALPs) \cite{26,27,28}, DPs are assumed to be produced in the
early Universe through certain production channels of inflationary
fluctuations, decays or annihilations of earlier fields \cite{26,27,28,
29,30,31}. Thus, depending on their production history, the resulting DPs
vector field may inherit specific properties that can be accessible at
present time in the intergalactic medium of our Galaxy. In particular, in
addition to the possibility that they can communicate with the visible
photons of the electromagnetic field through some portals, the DPs are
assumed to belong to the Weakly Interacting Slim Particles (WISPs) and have
a little mass via a specific mechanism, and hence can bring explanations for
some related phenomena, for instance the Hot DM (HDM), which does not appear
in the massless visible photons.

The principal goal of the present work is to deal with the massive DPs $%
A^{\prime }$ as a candidate for the HDM in the Universe. For that, we build
a model where the massive DPs consist of particles whose existence belongs
to a new symmetry beyond the SM. Concretely, we deal with a simple abelian
symmetry $U(1)_{X}^{\prime }$ extension of the SM and show how DPs $%
A^{\prime }$ can be readily lodged. Then, we study the breaking of such a
dark symmetry $U(1)_{D}^{\prime }$ and investigate the associated conserved
dark number $D$, that further we identify with the weak number $W$
associated with the the WISPs being a potential candidate for the HDM in the
Universe. Next, we disucss how this can be particularly helpful for the
breaking scale of this symmetry $U(1)_{W}^{\prime }$ and in supporting the
case for the massive DPs.

\section{Motivation for dark sector}

It is now known that the missing mass in the universe that we call DM is
about five times more abundant than the ordinary baryonic matter \cite{14,32}%
. Since DM cannot be integrated within the SM, its possible comprehension
demand the introduction of new degrees of freedom interacting weakly with
ordinary matter. The introduction of heavy particles $m_{X}\gtrsim M_{EW}$
above the energy reached we have reached so far is the most possible
direction. Also, DM could well be explained in terms of light states $%
m_{X}\ll 10^{2}GeV$ that are almost decoupled from the SM particles, so
called hidden sector. As long as its interaction with the SM particles is
weak, the mass scale of the hidden sector $m_{X}$ could be arbitrary. In
this approach, the new physics contribution can be represented as a
incorporation of two parts: an ultraviolet \footnote{%
Short distance.} term responsible for the new heavy degrees of freedom and
suppressed by the UV scale $\Lambda _{UV}$ , and an infrared \footnote{%
Long distance.} term \cite{33}. Thus, roughly, we can write

\begin{eqnarray}
L_{SM+NP} &=&L_{SM}+L_{NP}  \notag \\
&=&L_{SM}+L_{IR}+L_{UV}  \label{1} \\
&=&L_{SM}+F_{IR}\left( f_{SM},X_{IR}\right) +\frac{G_{UV}^{n}\left(
g_{SM},X_{UV}\right) }{\Lambda _{UV}^{m}}\text{. \ \ }  \notag
\end{eqnarray}%
where the functions $f_{SM}$ and $g_{SM}$ refer to terms involving products
of SM fields only, and the functions $F_{IR}$ and $G_{UV}$ refer to the new
terms of dimensions four and higher $d\geqslant 4$ involving products of SM
fields and the new field $X_{IR}$ and $X_{UV}$, respectively. The values of
the $n$ and $m$ numbers are such that the full dimension of the UV
contribution is four, i.e.,

\begin{equation}
n=4+m.  \label{2}
\end{equation}%
However, because no indication of new physics has been observed even with
the present highest energy reached at the colliders, the $UV$ conctribution
in the Lagrangian (\ref{1}) can be omitted and the full Lagrangian gets
simplified to

\begin{eqnarray}
L_{SM+NP} &=&L_{SM}+L_{IR}  \notag \\
&=&L_{SM}+F_{IR}\left( f_{SM},X_{IR}\right)  \label{3}
\end{eqnarray}%
which is the most general low-energy extension of the SM. It is so-called
hidden sector extension due to its extremely weak interaction with the
visible sector. The hidden sector could comprise rich states and
phenomenology, but seen that we are interested here in the low-energy
scenario which is extensively tested experimentally, we focus mostly on the
the minimal hidden boson model which has only three unknown parameters: the
mass of the new boson $m_{X_{IR}}$, its direct or indirect coupling $%
\varepsilon $ to the SM , and its decay branching fraction into hidden
sector final states $X_{IR}$ $\rightarrow Y_{IR}$ which is taken to be
either unity or zero depending on whether any invisible dark-sector final
states $Y_{IR}$ are kinematically allowed---that is, depending on whether $%
m_{X_{IR}}>2m_{Y_{IR}}$. Here we focus on the case where the field $X_{IR}$
is the lightest particle in the hidden sector and thus is stable. Therefore,
the corresponding parameter space of the minimal hidden boson model is

\begin{equation}
\wp _{_{SM+IR}}=\left( m_{X_{IR}},\varepsilon \right) .  \label{4}
\end{equation}%
In addition, we concentrate on $X_{IR}$ masses below the electroweak scale $%
m_{X_{IR}}\ll 10^{2}GeV$ which is the region accessible to accelerator-based
experiments, and where the $X_{IR}$ phenomenology is markedly different from
supersymmetry and other scenarios that extend the SM. In this view, the
existence of a new neutral vector particle

\begin{equation}
X_{IR}\equiv A^{\prime }  \label{5}
\end{equation}%
which has a non-vanishing coupling $\varepsilon $ to the SM is predicted.
Such a hidden photon $A^{\prime }$ could communicate with the visible sector
through a hidden messenger field $Z_{IR}$ but the communication can also be
achieved in different processes. Among the constraints on the arameters of
this dark photon scenario, the hidden photon must have a light $m_{\mathbf{A}%
^{\prime }}$\ mass and its coupling $\varepsilon $\ to the SM must be too
small so as to guarantee that its possible decay or annihilation processes
to SM sector must not affect the cosmic microwave background radiation. This
requirement will be considered in what follows through the direct
interaction of the hidden messenger field $Z_{IR}$\ with the SM sector.

\section{Dark Photon model}

\subsection{Dark Photon couplings and mass}

The economical model for the low-energy extenstion of the SM consists of
adding an abelian continuous symmetry; so that the group symmetry of the
extended model is

\begin{eqnarray}
G_{SM+X} &=&G_{SM}\times G_{X}  \notag \\
&=&SU\left( 3\right) _{C}\times SU\left( 2\right) _{I}\times U\left(
1\right) _{Y}\times U\left( 1\right) _{X}^{\prime }  \label{6}
\end{eqnarray}%
where the additional symmetry $U(1)_{X}^{\prime }$ is related to the
conservation of some quantum number $X$ to be determined later on. All SM
fields are assumed to be neutral under this additional symmetry so as

\begin{equation}
X\left( q_{i}\right) =X\left( e_{i}\right) =X\left( \upsilon _{i}\right) =0
\label{7}
\end{equation}%
where the SM fields $q_{i}$, $e_{i}$ and $\upsilon _{i}$ with $i$ being the
family number, refer to the SM quarks, electrons and left-handed neutrinos,
respectively. To allow for the simplest way of the breaking of this
additional symmetry, an extra scalar-like field $Z_{IR}$ neutral under $%
G_{SM}$ but charged under $U(1)_{X}^{\prime }$ \ which restores the
longitudinal polarization of the new vector boson $X_{IR}$ is introduced. We
call such a singlet scalar field hidden Higgs

\begin{equation}
Z_{IR}\equiv H^{\prime }  \label{8}
\end{equation}%
and whose the charges under the group symmetry of the extended model (\ref{6}%
) are

\begin{eqnarray}
C\left( H^{\prime }\right) &=&I\left( H^{\prime }\right) =Y\left( H^{\prime
}\right) =0  \label{9} \\
X\left( H^{\prime }\right) &\neq &0\text{.}  \label{10}
\end{eqnarray}%
Having this hidden Higgs $H^{\prime }$ added to the SM scalar content, the
most general renormalizable scalar potential reads

\begin{equation}
V_{SM+H^{\prime }}\left( H,H^{\prime }\right) =V_{SM^{\prime }}\left(
H\right) +\mu ^{2}\left\vert H^{\prime }\right\vert ^{2}+\lambda \left\vert
H^{\prime }\right\vert ^{4}+\lambda ^{\prime }\left\vert H\right\vert
^{2}\left\vert H^{\prime }\right\vert ^{2}  \label{11}
\end{equation}%
where $V_{SM^{\prime }}\left( H\right) $ is the standard Higgs potential of
the SM, $H=(h^{0},h^{-})^{T}$ is the SM Higgs doublet, and $\mu $, $\lambda $
and $\lambda ^{\prime }$ are real constants. More conveniently, it is useful
to detach the hidden Higgs field $H^{\prime }$ into its real and imaginary
parts by writing

\begin{equation}
H^{\prime }=\rho ^{\prime }e^{i\theta ^{\prime }}  \label{12}
\end{equation}%
where the real fields $\rho ^{\prime }$ and $\theta ^{\prime }$ stand for
the physical massive field and the massless Goldstone boson respectively.
With this at hand, the scalar potential (\ref{11}) becomes

\begin{equation}
V_{SM+H^{\prime }}\left( H,H^{\prime }\right) =V_{SM^{\prime }}\left(
H\right) +\mu ^{2}\rho ^{\prime 2}+\lambda \rho ^{\prime 4}+\lambda ^{\prime
}\left\vert H\right\vert ^{2}\rho ^{^{\prime }2}\text{.}  \label{13}
\end{equation}%
As it is known, the electroweak symmetry of the SM is surely broken by the
non-vanishing vacuum expectation value of the neutral field $h^{0}$ at the
energy scale order $\left\langle H\right\rangle \sim 10^{2}GeV$. For the $%
U(1)_{X}^{\prime }$ symmetry, the breaking is ensured if $-(%
%TCIMACRO{\U{b5}}%
%BeginExpansion
{\mu}%
%EndExpansion
^{2}+\lambda ^{\prime }\left\langle H\right\rangle ^{2})/2\lambda \rangle 0$%
, in such a case $\rho ^{\prime }$ develops a real vacuum expectation value
determined by

\begin{eqnarray}
\left\langle \rho ^{\prime }\right\rangle &=&\sqrt{\frac{m_{\rho ^{\prime
}}^{2}}{2\lambda ^{\prime }}}\text{,}  \label{14} \\
m_{\rho ^{\prime }} &=&\sqrt{-(%
%TCIMACRO{\U{b5}}%
%BeginExpansion
{\mu}%
%EndExpansion
^{2}+\lambda ^{\prime }\left\langle H\right\rangle ^{2})}\text{.}  \label{15}
\end{eqnarray}%
In this vision, the communication of the physical hidden Higgs $\rho
^{\prime }$ with the SM sector arises indirectly through the scalar portal $%
\lambda ^{\prime }|H|^{2}\rho ^{\prime 2}$ parameterizing the size of the
interaction between the SM Higgs boson and the hidden Higgs via the coupling
constant $\lambda ^{\prime }$. However, since the decay properties of the
standard Higgs boson with the SM rate expected would be impacted by this
interaction \cite{34,35}, such an interaction have to be very small $\lambda
^{\prime }\ll $ $1$. This corresponds to a weak mixing between the two
scalars $H$ and $\rho ^{\prime }$ generating a feeble mixing between the SM
weak $Z$-boson and the new $U(1)_{X}^{\prime }$ boson, and, subsequently, no
considerable impact on the decays of the $Z$ -boson.

Now, we can fo further and investigate the physical meaning of the supposed
symmetry $U(1)_{X}^{\prime }$. In fact, because there is no locus for a new
broken symmetry within the SM, it is then straightforward to think about a
symmetry associated with hidden particles feebly interacting with the SM,
but likely known to be plentiful in the Universe. We identify these
particles with massive Dark Photons DPs. Indeed, DPs are believed to be
abundant in the universe with a tiny mass scale. In this picture, one can
now assume that the preserved quantum number $X$ associated with the $%
U(1)^{\prime }$ symmetry introduced above (\ref{6}) is the Dark Number $D$:
the number of Dark Particles minus the number of their antiparticles. This is

\begin{equation}
X\equiv D=DP-\overline{DP}\text{.}  \label{16}
\end{equation}%
Using the identifification $U(1)_{X}^{\prime }\equiv U(1)_{D}^{\prime }$, we
consider the DP fields $A^{\prime }$, carrying, along with the dark Higgs
field $H^{\prime }$, a dark quantum number $D$ whilst all the SM fields are
again supposed to be neutral under $U(1)_{D}^{\prime }$ as

\begin{center}
\begin{tabular}{|l|l|l|l|l|}
\hline
Fields & $SM$ & $H$ & $H^{\prime }$ & $A^{\prime }$ \\ \hline
$DP$ & 0 & 0 & 2 & 1 \\ \hline
$\overline{DP}$ & 0 & 0 & 0 & 0 \\ \hline
$D$ & 0 & 0 & 2 & 1 \\ \hline
\end{tabular}
\end{center}

According to these field charges, among the new terms involving the dark
fields in the most general renormalizable extended lagrangian, we have

\begin{equation}
L_{SM+D}\supset \frac{1}{2}\varepsilon ^{2}\rho ^{\prime 2}A^{\prime 2}
\label{17}
\end{equation}%
where $\rho ^{\prime }$ is the physical dark Higgs and $\varepsilon $ is the 
$A^{\prime }-H^{\prime }$ coupling constant caracterizing the interaction
between the dark fields $\rho ^{\prime }$ and $A^{\prime }$. Owing to the
non-vanishing vacuum expectation value of the dark Higgs (\ref{14}), the
corresponding DP mass is

\begin{equation}
m_{A^{\prime }}=\frac{\varepsilon }{\sqrt{2}}\sqrt{\frac{m_{\rho }^{2}}{%
2\lambda ^{\prime }}}\text{.}  \label{18}
\end{equation}%
Pursuant to the cosmological observations which have directed to a
consistent model of the Universe where $\sim 85\%$ of matter is dark, i.e.,
non-baryonic \cite{14,32}, such massive abundant vector particles are a type
of candidates for the HDM and are notably quite favoured by many suggeted SM
extensions \cite{24,25}. These particles would inevitably contribute to the
HDM of the Universe. Thus, if these massive DPs $A^{\prime }$ actually form
the HDM, the dark number $D$ (\ref{16}) would then be nothing but the Weakly
Interacting Slim Number $W$: the number of WISPs minus the number of the
their antiparticles, being

\begin{equation}
D\equiv W=WISP-\overline{WISP}\text{.}  \label{19}
\end{equation}%
Again, using the identification $U(1)_{D}^{\prime }\equiv U(1)_{W}^{\prime }$%
, we can say that if these DPs really make up the HDM, they ought to have a
local mass density of that suposed in our vicinity to account for the
dynamics of our own galaxy. Concretely, they should be spread in a halo
bordering our galaxy with a typical relativistic velocity near the speed of
light $\sim c$, and would coherently scatter off nuclei in terrestrial
detectors \cite{36}. These DPs could be detected indirectly through their
self-annihilation $A^{\prime }\overline{A^{\prime }}\rightarrow x\overline{x}
$ results or directly by investigating their interaction in the detector via
their tiny shocks with its nuclei with a typical \ kinetic energy $%
K_{A^{\prime }}$ of tens of $KeV$'s, according to the (CDMS II) data \cite%
{37}. Using this, the DP mass is, roughly, at most

\begin{equation}
m_{A^{\prime }}\sim \frac{K_{A^{\prime }}}{c^{2}}\leq 10^{2}KeV  \label{20}
\end{equation}%
which in turn allows to approach the related scales of the model. Indeed,
for coupling constants taken at most $\lambda ^{\prime }\sim \varepsilon $ $%
\sim \vartheta \left( 1\right) $, we get for the physical dark Higgs mass
and its vacuum expectation value

\begin{eqnarray}
m_{\rho ^{\prime }} &\sim &\frac{2\sqrt{\lambda ^{\prime }}m_{A^{\prime }}}{%
\varepsilon }\geq 10^{2}KeV  \label{21} \\
\left\langle \rho ^{\prime }\right\rangle &\sim &\frac{2m_{A^{\prime }}}{%
\varepsilon }\geq 10^{2}KeV  \label{22}
\end{eqnarray}%
where we see that all the DP involved scales (\ref{14}), (\ref{15}) and (\ref%
{18}) appear to lie under the well-probed electroweak scale $\sim 10^{2}GeV$.

\subsection{Dark sector decays and productions}

In the present model, the DPs $A^{\prime }$\ do not interact directly with
the visible sector; but they interact indirectly with the SM via the dark
Higgs $H^{\prime }$\ as discussed previously. DPs are assumed to stable
because of their light mass and the conservation of the WISP number
associated to the new symmetry $U(1)_{D}^{\prime }$. Since the DP is
electrically neutral and is its own anti-particle $A^{\prime }\equiv 
\overline{A^{\prime }}$\ and ccording to its mass range (\ref{20}), a DP
pair in the early Universe can self-annihilate to produce light SM
particles, mainely photons $A$\ and lighter neutrinos

\begin{equation}
A^{\prime }A^{\prime }\rightarrow \gamma \gamma ,\text{ }\upsilon
_{e}\upsilon _{e}\text{.}  \label{23}
\end{equation}%
However, as the Universe expands and DPs annihilate without being produced
back from lighter particles, their density drops significantly until they
are so diluted that they stop interacting with each other. This appeasing
process leaves a constant population of DPs whose density decreases with the
expanding space.

The production as well as the abundance of the DPs can now be discussed. In
the present model, the DPs can be produced by dark Higgs via annihilation
processes according to some specific modes, depending on the relative mass
between $A^{\prime }$ and $\rho ^{\prime }$ particles. Indeed, because we
have\ $m_{\rho ^{\prime }}>m_{A^{\prime }}$, DPs can be produced via the
dark Higgs annihilation process like

\begin{equation}
\rho ^{\prime }\rho ^{\prime }\rightarrow A^{\prime }A^{\prime }  \label{24}
\end{equation}
as shown in the following figure

\begin{equation*}
\FRAME{itbpF}{3.3157in}{2.6593in}{0in}{}{}{Figure}{\special{language
"Scientific Word";type "GRAPHIC";maintain-aspect-ratio TRUE;display
"USEDEF";valid_file "T";width 3.3157in;height 2.6593in;depth
0in;original-width 3.2707in;original-height 2.6178in;cropleft "0";croptop
"1";cropright "1";cropbottom "0";tempfilename
'RLPK5C00.wmf';tempfile-properties "XPR";}}
\end{equation*}

\begin{center}
Fig.: Production of a dark photon via dark Higgs annihilation.
\end{center}

The corresponding rate of the this process scales as

\begin{equation}
\left\langle \sigma \upsilon \right\rangle _{\rho ^{\prime }\rho ^{\prime
}\rightarrow AA^{\prime }}\sim \frac{\varepsilon ^{4}}{m_{\rho ^{\prime
}}^{2}}\leqslant 10^{-10}eV^{-2}  \label{25}
\end{equation}%
which only relies on the dark coupling $\varepsilon $ and on the mass of the
dark Higgs $m_{\rho ^{\prime }}$ but not on the visible-dark Higgs $%
H-H^{\prime }$ mixing parameter $\lambda ^{\prime }$ -- which means that the
process would be difficult to be detected by a particle physics experiment.
With only the degrees of freedom in the model and based on (\ref{18}), (\ref%
{21}), (\ref{22}) and (\ref{25}) and taking into account the dilution of the
DPs, straightforward calculations lead the resulting DPs abundance which can
be read, roughly, as

\textbf{\bigskip }%
\begin{equation}
\Omega _{A^{\prime }}\sim 0.2\left( \frac{\varepsilon }{10^{-8}}\right) ^{3}
\label{26}
\end{equation}%
where on can see that for a value $\varepsilon \sim 10^{-3}$\ of the DP
coupling the right abundance of the DM can be obtained.\textbf{\ }

Although many scenarios and production mechanisms of DP models have been
discussed in the literature, DP with a tiny mass below $10^{2}KeV$ scale is
a motivated candidate for the HDM and its existence remains well motivated
from particle physics.

\section{Conclusion}

WISP candiates for the HDM appear to be a generic possibility, as rather
minimal model building choices lead to viable WISPs interacting with
ordinary matter through metastable mediators. They are now the target of a
growing number and type of experimental searches that are complementary to
new physics searches at colliders.

In this work, we have investigated a possible WISP candidate for the HDM in
an infrared extension of the SM consisting of an extra symmetry $%
U(1)_{D}^{\prime }$ associated with the new DP $A^{\prime }$. The model
contained, in addition to the new boson $A^{\prime }$, a dark Higgs field $%
H^{\prime }$ responsible for the spontaneous breaking of the dark symmetry $%
U(1)_{D}^{\prime }$. The corresponding conserved quantum number $D$ has been
associated with the massive DPs $A^{\prime }$. Under this coincidence of a
conserved dark quantum number $D$, the dark Higgs $H^{\prime }$ couples to
the DP $A^{\prime }$ whose mass is generated after spontaneous breaking of
the dark symmetry $U(1)_{D}^{\prime }$ by the non-vanishing dark Higgs vev $%
\left\langle \rho ^{\prime }\right\rangle $. Then, we went even deeper and
considered the possibility where the dark symmetry $U(1)_{D}^{\prime }$
corresponds to the HDM particles for which the conservation of the dark
number $D$ was identified with the WISP number $W$. With this assessment,
the DPs should have a typical speed and interaction with terrestrial
detectors according to DM search experiment data which permitted, after
investigating the DPs $A^{\prime }$ kinetic energy, to approximate the
involved mass scales of the model, with a DP mass $\leq 10^{2}KeV$, an
energy scale $\geq 10^{2}KeV$ for the dark symmetry $U(1)_{W}^{\prime }$
breaking as well as for the associated physical dark Higgs $\rho ^{\prime }$.

The DP physics still attract much interest due to the possibilities it
provides for the explanation of several phenomena at the same time: the
possible indications for DM scattering signals in high purity experiments,
the dark matter annihilation mechanism, and more...

\begin{acknowledgement}
: S.E.E. would like to thanks his family for support.
\end{acknowledgement}

\end{document}